\documentclass[aps, prb, reprint, floatfix, superscriptaddress]{revtex4-2}

\usepackage{amsmath}
\usepackage{graphicx}
\usepackage{color}
\usepackage{tabularx}
\usepackage{xr}
\usepackage{floatrow}
\usepackage{xcolor}
\usepackage{upgreek}

\usepackage{moreverb} 

\usepackage{verbatim}

\makeatletter

\begin{document}

\title{Statistical Properties of a Fluctuation-Driven Nanomechanical Duffing Resonator}

\author{M. Ma}
\affiliation{Department of Mechanical Engineering, Division of Materials Science and Engineering, and the Photonics Center, Boston University, Boston, Massachusetts 02215, USA \looseness=-1}

 \author{N. Welles}
 \affiliation{Department of Mechanical Engineering, Virginia Tech, Blacksburg, Virginia 24061, USA \looseness=-1}

\author{J. Barbish}
 \affiliation{Department of Engineering, James Madison University, Harrisonburg, Virginia 22807, USA \looseness=-1}
 

\author{I. I. Kaya}
\affiliation{SUNUM, Nanotechnology Research and Application Center, Sabanci University, Istanbul, 34956, Turkey \looseness=-1}
\affiliation{Faculty of Engineering and Natural Sciences, Sabanci University, Istanbul, 34956, Turkey \looseness=-1}


\author{M. S. Hanay}
\affiliation{Department of Mechanical Engineering, Bilkent University, Ankara, 06800, Turkey \looseness=-1}
\affiliation{National Nanotechnology Research Center (UNAM), Bilkent University, Ankara, 06800, Turkey \looseness=-1}


\author{O. Svitelskiy}
\affiliation{Department of Physics, Gordon College, Wenham, Massachusetts 01984, USA \looseness=-1}

\author{M. R. Paul}
\email{mrp@vt.edu}
\affiliation{Department of Mechanical Engineering, Virginia Tech, Blacksburg, Virginia 24061, USA \looseness=-1}

\author{K. L. Ekinci}
\email{ekinci@bu.edu}
\affiliation{Department of Mechanical Engineering, Division of Materials Science and Engineering, and the Photonics Center, Boston University, Boston, Massachusetts 02215, USA \looseness=-1}

\date{\today}

\begin{abstract}

We investigate the fluctuating nonlinear dynamics of multiple modes of a nanomechanical doubly clamped beam resonator. Each mode is driven by Gaussian force noise centered around the resonance; the mode response is monitored while the force magnitude is increased, inducing a transition from harmonic to the nonlinear Duffing regime. To characterize the dynamics, we introduce an effective temperature based on the mode fluctuation amplitude. As the nonlinearity becomes prominent at large amplitudes, we observe a reduction in the effective thermal energy and a crossover in response statistics from Gaussian to platykurtic, consistent with the dynamics expected in a Duffing potential.
\end{abstract}


\maketitle

Certain phenomena in macroscopic systems can be formulated effectively in terms of fluctuating drive forces \cite{dykman1985spectral,dykman1980time}. These stochastic forces typically enter the dynamical equations of motion through statistical averaging over microscopic degrees of freedom and capture the coupling of the system to other internal or external degrees of freedom \cite{lee2017fluctuation}.  For instance, our understanding of thermal equilibrium of a simple macroscopic system, such as a Brownian particle or an electrical resistor, can be boiled down to the presence of a generalized fluctuating force from the thermal bath; a  Langevin equation then describes the system-bath interactions in terms of a fluctuating mechanical force or an electrical charge.  In active non-equilibrium phenomena, such as transport in biological systems or turbulent flows, the stochastic drive force provides the energy to sustain fluctuations and break detailed balance. Whether equilibrium or non-equilibrium, such forcing typically carries signatures of the underlying dynamics as well as the characteristics of the physical system and introduces statistics into the observables. Despite efforts to generalize \cite{battle2016broken, wensink2012meso, mizuno2007nonequilibrium},  principles for obtaining these fluctuating forces and for interpreting the response to these forces remain incomplete. 

In a canonical equilibrium system, such as the above-mentioned Brownian particle in a harmonic potential, the fluctuating force is delta-correlated in time (white in the frequency spectrum) with a Gaussian probability distribution function (PDF) and tuned to satisfy the fluctuation–dissipation theorem \cite{arcizet2006high,gloppe2014bidimensional,paolino2009direct,saulson1990thermal}. Consequently, the particle exhibits position fluctuations at the root-mean-squared  (rms) amplitude predicted by classical equipartition, with a Gaussian PDF. The response spectrum in this case, assuming small dissipation, is sharply peaked at the natural frequency, reflecting linear dynamics and the absence of energy transfer across (frequency) scales. By contrast, in homogeneous isotropic turbulence, a similar white Gaussian force noise produces non-Gaussian velocity statistics due to the nonlinear Navier–Stokes dynamics, and the input energy cascades through scales, generating a broad, power-law spectrum  \cite{pope2001turbulent}. The goal of this paper is to examine how introducing nonlinearity changes the fluctuation-driven response of a canonical system, a Duffing resonator, which combines a harmonic term with a weak quartic nonlinearity in its potential \cite{strogatz2024nonlinear,goldobin2005synchronization,nayfeh2008nonlinear, lifshitz2008nonlinear,schmid2016fundamentals}. We drive the Duffing resonator above thermal equilibrium with a thermal-like force noise. An analysis of the response in terms of an effective temperature reveals changes in the response spectrum and statistics due to nonlinearity. 

\begin{figure*}
    \includegraphics{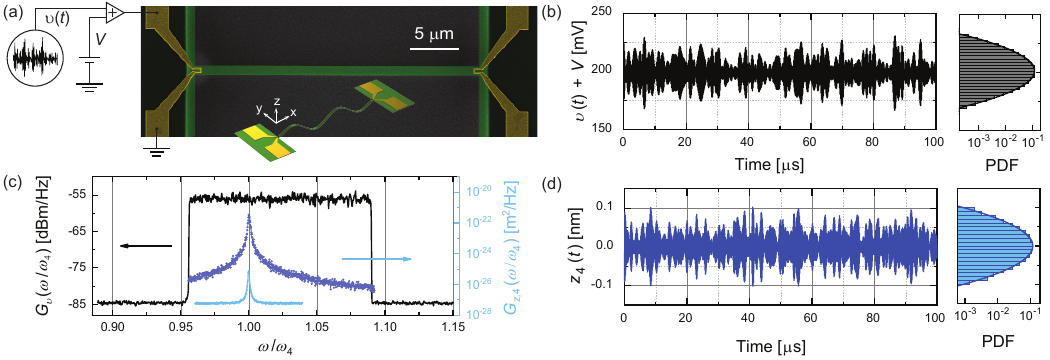}
    \caption{(a) SEM of the doubly clamped nanomechanical beam resonator and the electrical drive schematic. The drive is a band-limited noise $v(t)$ with a DC offset $V$. The inset shows the $n=4$ eigenmode with vibrations along the $z$-axis. (b) The noisy drive voltage in the time domain (left) and its PDF (right). The continuous line is a Gaussian. (c) Typical PSDs of the drive and linear response of mode $n=4$. The light blue curve is the response at $v(t)=0$ and $V=200 ~\rm{mV}$; the black curve is the drive at $v(t)$ with an rms value of 9.4 mV and $V=200 ~\rm{mV}$; the dark blue curve is the response at this drive. \textcolor{black}{For both the thermal (light blue) and the noise driven (dark blue) responses, $Q_4 \sim10^3$.} (d) The linear response in the time domain (left) also has a Gaussian PDF (right).}
    \label{Figure_1}
\end{figure*}

Fig. \ref{Figure_1}(a) shows a false-colored scanning electron micrograph (SEM) of our resonator, a suspended doubly clamped SiN beam with dimensions $l\times w\times h = 30~\rm{\upmu} m \times 900 ~\rm{nm}\times100~\rm{nm}$. The beam is driven by one of the two electrothermal transducers on its anchors to excite vibrations of the first $4$ eigenmodes in the $z$ direction \cite{ma2023electrothermal,bargatin2007efficient}. The inset in Fig.  \ref{Figure_1}(a) shows the fourth eigenmode.  Fig.  \ref{Figure_1}(b) shows a typical drive voltage in the time domain along with its PDF. Here, the electrothermal voltage is the sum of a DC offset $V=200 ~\rm{mV}$ and a narrow-band noise voltage $v(t)$ with an rms value of $9.4 ~\rm{mV}$, where $v(t)$ has a Gaussian PDF.

The measured power spectral density (PSD) of $v(t)$, $G_v({\omega \over \omega_4})$, where the angular frequency $\omega$ is normalized by the fourth (angular) eigenfrequency $\omega_4$, is the black data trace in Fig.  \ref{Figure_1}(c). It is a rectangular function of frequency with a bandwidth of $\Delta f \approx 1~\rm MHz$ centered around  $\omega_4/2\pi \approx 7.6~\rm MHz$. The drive force is proportional to $(V+v)^2$ \cite{ma2023electrothermal}. Since $V \gg v$, the relevant drive force noise is proportional to $2vV$ and has a Gaussian PDF \cite{Allemeier2025Multistability}. In our experiments, we fix $V$ and gradually increase $v(t)$  while maintaining the condition  $V \gg v$ and $\Delta f \gg \omega_n/2\pi Q_n$, where $Q_n$ is the modal quality factor. \textcolor{black}{This condition ensures that the force noise correlation time of $\sim 10^{-6}~\rm s$,  is much shorter than the typical modal relaxation time of $\sim 10^{-4} ~\rm s$, so that the drive can be assumed white for each mode.} The detailed process for generating the drive noise, along with its characteristic properties, is provided in the Supplementary Material.

The fluctuations $z_n(t)$ of each mode are detected using a Michelson interferometer at an antinode near the beam's center, in both the frequency and the time domain. The light blue curve in Fig.  \ref{Figure_1}(c) is the PSD of $z_4(t)$, $G_{z,4}({\omega \over \omega_4})$,  in the presence of the DC offset only, \textit{i.e.,} $V=200 ~\rm{mV}$ and $v(t) = 0$. The dark blue curve in Fig.  \ref{Figure_1}(c) shows $G_{z,4}({\omega \over \omega_4})$  in the presence of the noisy drive with $G_v({\omega\over2\pi})$ (black curve).  $G_{z,4}({\omega\over2\pi})$ is still in the linear regime. In the time domain, $z_4(t)$ follows the typical response of a linear oscillator with small damping, revealing a Gaussian PDF [Fig.  \ref{Figure_1}(d)].


Our experimental observations arise from the intrinsic geometric nonlinearity in doubly clamped beams when the beam stretches at high vibrational amplitudes \cite{ma2024mode, kozinsky2007basins, postma2005dynamic, matheny2013nonlinear, Allemeier2025Multistability}. The dynamical equation of mode $n$ can be written as
\begin{equation}\label{eqn:duffing_v2}
    {\ddot{z}_n} + {\frac{{\omega_n}}{{Q_n}}}{\dot{z}_n} +  \left({{\omega_n}^2} + {\alpha_n}{z_n}^2\right) {z_n}= \frac{\xi(t)}{m_n}.
\end{equation}
Here, $\alpha_n$ is the  Duffing constant; $m_n = k_n/{\omega_n}^2$ is the modal mass, where $k_n$ is the linear spring constant and  $\omega_n$ is the eigenfrequency. Dots indicate time derivatives. All the relevant parameters for the first $4$ modes are shown in Table \ref{table1:nominals}. The stochastic drive force $\xi(t)$ arises from the band-limited drive voltage as described above. In our experiments, we gradually increase $\xi(t)$ by increasing $v(t)$, thereby driving a mode into its Duffing regime.  We note that the Joule heating from the drive increases the thermodynamic temperature of the beam \cite{ma2023electrothermal} and changes the mechanical parameters. We estimate that across modes, the maximum thermodynamic temperature increase (at the maximum drive) is $100~\rm K$. 
\textcolor{black}{In our theoretical calculations, we use the perturbed $\omega_n$ and $k_n$ values, listed in parentheses in Table \ref{table1:nominals}. Changes in $Q_n$ are less consequential, and $\alpha_n$ is only weakly dependent on the thermodynamic temperature \cite{ma2024mode}, justifying the use of their unperturbed values.} More details on this can be found in the Supplementary Material.

\begin{table}
\caption{Mechanical parameters for the first 4 modes. The values in parentheses indicate the maximum changes due to  Joule heating.}\label{table1:nominals}
\renewcommand{\arraystretch}{1.1}
\newcolumntype{Y}{>{\centering\arraybackslash}X}
\begin{tabularx}{\textwidth}{Y|YYYY}
 \hline
 Mode $n$& $\omega_n/2\pi$ & $Q_n$ & $k_n$ & $\alpha_n/4\pi^2$ \\
 & $(\rm{MHz})$ &  & $(\rm N/m)$ & $(\rm MHz^2/nm^2)$\\
 \hline
1 & 2.37 (1.74) & $8.7\times 10^3 $&0.91 (0.78)& $6.65 \times 10^{-5}$\\
2 & 4.90 (2.75)& $6.1\times 10^3$  &3.44 (2.50)& $1.01 \times 10^{-3}$\\
3 & 8.01 (5.37)& $3.2\times 10^3$  & 10.0 (8.18)& $5.14 \times 10^{-3}$\\
4 & 10.5 (7.66)& $1.8\times 10^3$    &19.2 (16.3)& $1.63 \times 10^{-2}$\\

\end{tabularx}
\end{table}

\begin{figure}
\includegraphics{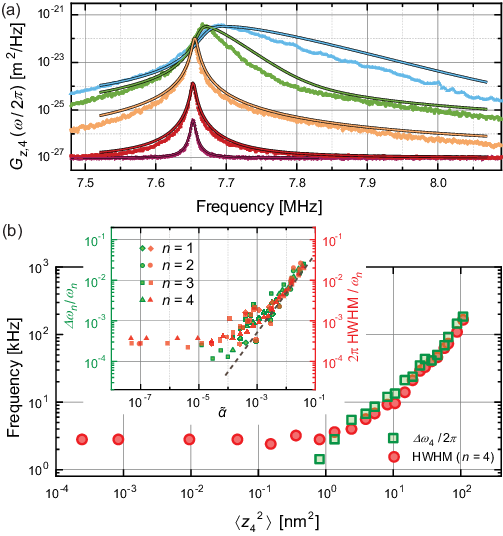}
{\caption{(a)  PSDs at different noise drive levels for mode $n=4$. Continuous curves are from theory \cite{dykman1980fluctuations}.  (b) Nonlinear frequency shift (green squares) and the half-width at half maximum (red circles) as a function of $\langle {z_4}^2\rangle$ on a double logarithmic axes. The inset shows the same data nondimensionalized by  $\omega_n$ for all modes as a function of the dimensionless Duffing constant $\tilde \alpha$. The dashed line is proportional to  ${\tilde \alpha}^2$.}
\label{Figure_2}}
\end{figure}

In Fig.  \ref{Figure_2}(a), we show several representative $G_{z,4}({\omega\over 2\pi})$  with increasing $v(t)$, where the lowest curve corresponds to the baseline response at $v(t) = 0$. The continuous curves are found from the Duffing theory \cite{dykman1971classical,dykman1980time, welles2024mode} with a small offset to account for the measurement noise (see Supplementary Material). The two curves with the largest amplitude in Fig. \ref{Figure_2}(a) are in the nonlinear regime. Here, we observe the broadening of the peak and an upward shift in the peak frequency --- typical features of a stiffening Duffing nonlinearity. \textcolor{black}{The experimental data deviate slightly from theory at larger drives, with the possible reasons discussed in the Supplementary Material.}

In Fig.  \ref{Figure_2}(b), we show the nonlinear frequency shift $\Delta \omega_{4}/2\pi$  and the half-width at half maximum (HWHM) of the same mode ($n=4$) as a function of the mean-squared response amplitude $\langle {z_4}^2\rangle$. The nonlinear frequency shift is found from PSDs, such as the ones shown in Fig.  \ref{Figure_2}(a), as $\Delta \omega_{4}/2\pi \approx (\omega_{p,4}-\omega_4)/2\pi$, where $\omega_{p,4}/2\pi$ is the shifted peak frequency. The HWHM (red circles) remains constant for $\langle {z_4}^2\rangle \lesssim 1~\rm{nm^2}$; this is the linear regime, where  ${\rm HWHM} \approx \frac{\omega_4}{4 \pi Q_4}$. 
For $\langle {z_4}^2\rangle\gtrsim 1~\rm{nm^2}$, displacement fluctuations are transduced into frequency fluctuations due to the nonlinear frequency-amplitude coupling \cite{dykman1980fluctuations}. As a result, the response PSD asymmetrically extends toward higher frequencies and becomes an average of amplitude-shifted linear PSDs that are weighted by the Gibbs distribution \cite{gieseler2013thermal}. At this limit, the HWHM is also determined by the nonlinear frequency shift. The inset of Fig.  \ref{Figure_2}(b) shows similar data for all the modes plotted in dimensionless units normalized by the eigenfrequency. The  $x$-axis is plotted in units of a dimensionless Duffing constant $\tilde{\alpha}$ (also called the frequency straggling parameter \cite{dykman1980time}). More details on how $\tilde{\alpha}$ is determined for this system are presented below. 

The  data  in Fig.  \ref{Figure_2}(b) shows that $\Delta \omega_n \propto \langle {z_4}^2\rangle$, also known as the Duffing backbone curve \cite{kozinsky2006tuning, ma2024mode, Allemeier2025Multistability}. Since our system is  stochastically driven, the backbone curve follows \cite{dykman1980time}
\begin{equation}\label{eqn:backbone}
   \frac{\Delta \omega_n}{\omega_n} \approx \frac{3}{8} \frac{\alpha_n}{{\omega_n}^2} \langle {z_n}^2\rangle.
\end{equation}
Eq. (\ref{eqn:backbone}) is approximate because the variance $\langle {z_n}^2\rangle$ of the Duffing resonator starts to deviate from that of a linear resonator, \textit{i.e.,} ${k_B T_{\text{eff}}/ k_n}$, when $\tilde{\alpha}$ becomes comparable to unity \cite{dykman1980time}, as we discuss below. Using Eq. (\ref{eqn:backbone}), we find the Duffing constant $\alpha_n$ in Table \ref{table1:nominals} from experimental measurements of  $\Delta \omega_n$ and $\langle {z_n}^2\rangle$.  Our values of $\alpha_n$ agree with the backbone curve obtained from deterministic drive experiments. 

\begin{figure}
    \includegraphics{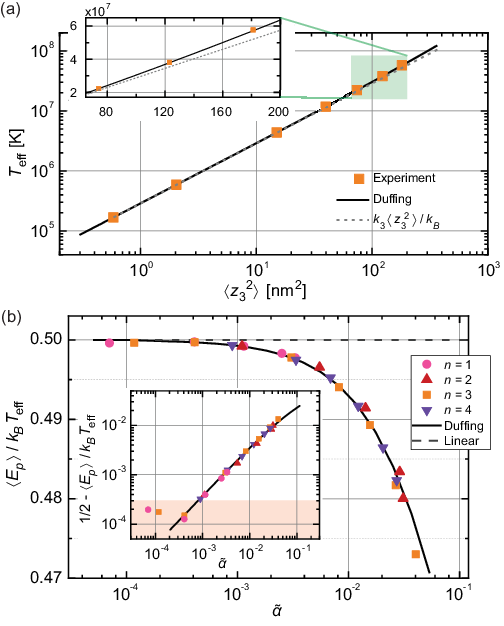}
    \caption {(a)  $T_{\text{eff}}$   found from $\langle {z_3}^{2}\rangle$ for mode $n=3$. Symbols correspond to the experimental values. The inset shows the highlighted region in linear axes, revealing a deviation from $T_{\text{eff}}$ of a linear oscillator (dotted line). (b)  ${\langle E_p\rangle}/k_B T_{\text{eff}}$ as a function of $\tilde{\alpha}$. Experimental values are shown as colored symbols. The dashed line is the harmonic oscillator ${\langle E_p\rangle}/k_B T_{\text{eff}}$.  The inset shows $\frac{1}{2}-\frac{\langle E_p\rangle}{(k_BT_{\text{eff}})}$ on double logarithmic axes. The shaded area is the experimental uncertainty.}
    \label{Figure_3}
\end{figure}

We now introduce an effective temperature $T_{\text{eff}}$ for this system that is actively driven above thermal equilibrium \textcolor{black}{\cite{martinez2013effective,mestres2014realization,ricci2017optically,yang2020phonon}}. This $T_{\text{eff}}$ value is found by matching the measured $\langle {z_n}^2\rangle$ to the theoretical  $\langle {z_n}^2\rangle$ for a Duffing oscillator in thermal equilibrium at $T_{\text{eff}}$. For a Duffing oscillator \cite{roberts2003random}
\begin{equation}\label{eqn:variance-theory}
    \langle{z_n}^2\rangle = \frac{k_B T_{\text{eff}}}{k_n} \sqrt[4]{\frac{2\pi^2}{\tilde{\alpha}}} \mathcal{D}_{-3/2} \left(\sqrt{\frac{1}{2\tilde{\alpha}}} \right)\mathcal{K}_{1/4}^{-1}\left(\frac{1}{8\tilde{\alpha}} \right).
\end{equation}
In Eq. (\ref{eqn:variance-theory}), $\mathcal{K}_l(a)$ and $\mathcal{D}_l(a)$ are the modified Bessel function of the second kind and the parabolic cylinder function of order $l$  and argument $a$ ~\cite{roberts2003random}. The Duffing constant $\alpha_n$ is subsumed in a dimensionless Duffing constant defined as 
\begin{equation}\label{eqn:alpha_tilde_theory}
    {\tilde\alpha} = \frac{\alpha_n}{{\omega_n}^2} \frac{k_B T_{\text{eff}}}{k_n}.
\end{equation}
Eq. (\ref{eqn:variance-theory}) allows us to compute the $T_{\text{eff}}$ value for each  $\langle {z_n}^2\rangle$  based on experimental $k_n$, $\omega_n$, and $\alpha_n$ values. \textcolor{black}{Since $T_{\text{eff}}$ is determined from $\langle {z_n}^2\rangle$, the uncertainty in the thermodynamic temperature does not propagate to $T_{\text{eff}}$.} Fig.  \ref{Figure_3}(a) shows calculated $T_{\text{eff}}$ as a function of $\langle {z_3}^2\rangle$ in a double logarithmic plot for mode $n=3$, with the squares corresponding to measured $\langle {z_n}^2\rangle$ from PSDs at different drives. In the inset, we focus on a portion of the data plotted on a linear scale, showing a slight deviation of the Duffing $T_{\text{eff}}$  from that of the linear oscillator, \textit{i.e.,} ${k_3 \langle{z_3}^2\rangle/ k_B}$ (dotted line). We note that the highest value of $\tilde{\alpha}$ attained in our experiments is $\approx 4\times 10^{-2}$ at $T_{\text{eff}}\approx 5.8\times 10^7~\rm K$ for mode $n=3$, where $\langle {z_3}^2\rangle$ of the Duffing oscillator is $\approx 90\%$ of that of a harmonic oscillator. Similar curves for each mode provide our effective temperature calibration.

After finding the $T_{\text{eff}}$ values, we turn to the  potential energy of our Duffing system, 
${E_p} = \frac{1}{2} {k_n {z_n}^{2}} + \frac{1}{4} \frac{\alpha_n k_n}{{\omega_n}^{2}}  {z_n}^{4}$.
We calculate ${\langle E_p\rangle}$ of the system for each measured $z_n(t)$ data set based on experimental  $k_n$, $\alpha_n$ and $\omega_n$ values, where $\langle {z_n}^{2}\rangle$ and $\langle {z_n}^{4}\rangle$ values are computed from $z_n(t)$ as described in the Supplementary Material. Fig.  \ref{Figure_3}(b) shows the experimental ${\langle E_p\rangle}$  plotted in dimensionless form,  ${\langle E_p\rangle}/k_B T_{\text{eff}}$, as a function of $\tilde \alpha$, which now corresponds to the dimensionless effective temperature of the Duffing system.   As $T_{\text {eff}}$ is increased and the mode samples the higher-order nonlinear potential more frequently, the mean energy decreases from $\frac{1}{2} k_B T_{\text{eff}}$ (dashed line). The inset in Fig. \ref{Figure_3}(b) shows this deviation, namely, $\frac{1}{2}-{\langle E_p\rangle \over {k_B T_{\text{eff}}}}$, 
as a function of $\tilde{\alpha}$ on a double-logarithmic plot, along with the experimental uncertainty. 


The continuous curve in Fig.  \ref{Figure_3}(b) is the theoretical  $\langle E_p\rangle$  derived from the partition function of a Duffing oscillator \cite{barbish2023dynamics}:
\begin{multline}\label{eqn:John-energy-theory}
    \frac{\langle E_p\rangle}{k_B T_{\text{eff}}} = \frac{1}{8\tilde{\alpha}}  \left( \frac{\mathcal{K}_{-3/4}(\frac{1}{8\tilde{\alpha}}) - 2\mathcal{K}_{1/4}(\frac{1}{8\tilde{\alpha}}) + \mathcal{K}_{5/4}(\frac{1}{8\tilde{\alpha}})}{2\mathcal{K}_{1/4}(\frac{1}{8\tilde{\alpha}})}\right).
\end{multline}
Even though the deviation from a harmonic oscillator is small, our experiment resolves it and achieves excellent agreement with theory.

\begin{figure}
    \includegraphics{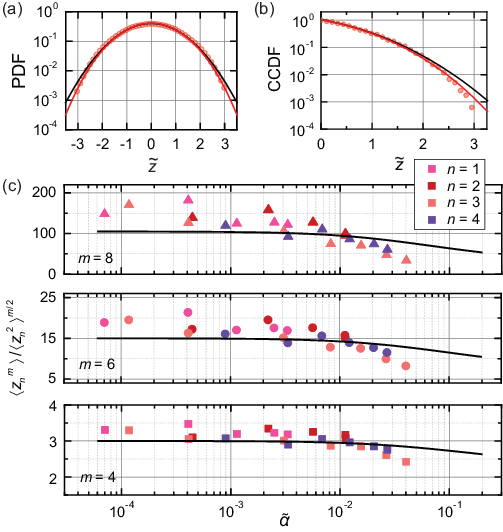}
    \caption{(a,b) Semilogarithmic plot showing the experimental PDF (a) and CCDF (b) of a nonlinear response (symbols), $\tilde{\alpha} \approx 3\times10^{-2}$, which matches the theoretical PDF curve (red), and exhibits platykurtic features compared to the Gaussian curve (black). (c) Semilogarithmic plot of experimental standardized $m^{th}$ order moments (colored symbols) and theory curves as a function of $\tilde{\alpha}$ for all modes: $m=8$ (top), $m=6$ (middle), and $m=4$ (bottom).}  
    \label{Figure_4}
\end{figure}

For more insight into the nonlinear stochastic dynamics, we turn to the statistical analyses of $z_n(t)$. We plot a Gaussian PDF and a typical nonlinear PDF of $z_n(t)$ with $\tilde \alpha \approx 3\times 10^{-2}$  in Fig.  \ref{Figure_4}(a).   Fig.  \ref{Figure_4}(b) shows the Complementary Cumulative Distribution Function (CCDF), which measures the probability that $\tilde{z}$ takes on a value greater than a given value. The $x-$axes are plotted in dimensionless displacement units $\tilde{z} = z_n/\sqrt{k_BT_\text{eff}/k_n}$. Both in Fig. \ref{Figure_4}(a) and (b), the nonlinear experimental data deviates from a linear curve (Gaussian, black) due to a contribution from the Duffing potential. The theoretical PDF of a Duffing resonator (red) is obtained from the Boltzmann distribution and using the experimental $\tilde{\alpha}$ as $P(\tilde{z}, \tilde\alpha)\approx {1\over N(\tilde \alpha)}e^{-\frac{1}{2}{\tilde{z}}^2 - \frac{1}{4}\tilde{\alpha}{\tilde{z}}^4}$, where the energy is expressed in dimensionless units and $N(\tilde \alpha)$ is a normalization constant.   Based on the length of our shortest data set and the correlation time in our experiments, which is approximately the ringdown time of a mode, $2\pi Q_n/\omega_n$, we estimate that our data are accurate up to $\sim 3 |\tilde{z}|$, as discussed in detail in the Supplementary Material. 


To characterize the subtle changes in the tails of our PDF curves, we calculate the moments of order $m$, $ \langle{z_n}^{m}\rangle $, numerically from the measured $z_n(t)$. The odd moments vanish due to the  symmetry; for even $m$, the  standardized moments, ${\langle{z_n}^{m}\rangle}/{\langle{z_n}^{2}\rangle^{m/2}}$, are presented in Fig.  \ref{Figure_4}(c) as a function of $\tilde \alpha$ for all measured modes. The continuous curves correspond to the theoretical values of the standardized moments of a Duffing oscillator (see Supplementary Material). In the limit of a purely quadratic (quartic) potential, the numerical values for $m=4,6,$ and $8$ are 3 (2.188), 15 (6.565), and 105 (23.95), respectively \cite{dykman1980time}. Our experimental data  agree with a Gaussian distribution for $\tilde{\alpha} \lesssim 10^{-2}$ and start to decrease for $\tilde{\alpha} \gtrsim 10^{-2}$.  The noise in the moment values is again attributed to the finite time data and the correlation time in our experiments \cite{gieseler2013thermal}. The error increases further for higher modes, given the increased sensitivity to rare fluctuations. Our estimated rms error with respect to the theory is $\approx 4\%, 11\%, \text{and}~ 20\%$ for $m=4, 6,\text{and}~8$, respectively.



In this work, we have studied the fluctuating nonlinear response of $4$ modes of a nanomechanical resonator. We have found quantitative agreement between experiments and theory, indicating that fluctuations provide a direct probe of the underlying nonlinear dynamics and energy landscape. Our approach and results may apply to other nonlinear systems, including granular and turbulent flows, as well as quantum and biological systems.  Beyond non-equilibrium systems, our approach also provides insight into systems that exhibit nonlinearities under thermal equilibrium \cite{gieseler2013thermal}.

\section*{DATA AVAILABILITY}
The data that support the findings of this study are available from the corresponding authors upon reasonable request.

\begin{acknowledgments}
\noindent K. L. Ekinci and M. Ma acknowledge support from the National Science Foundation (NSF) Grant Nos. CMMI-1934271, CMMI-2001403 and CMMI-2337507.  M. R. Paul, N. W. Welles, and J. Barbish acknowledge support from the NSF Grant No. CMMI-2001559. O. Svitelskiy acknowledges support from the NSF Grant No. CMMI-1934370 and CMMI-2337506. The authors acknowledge the use of the Boston University Photonics Center shared facilities. This work was performed in part using the shared laboratories and instrumentation supported by the Boston University Photonics Center. The authors  thank H. Gress for the SEM image of the NEMS device.
\end{acknowledgments}

\bibliographystyle{apsrev4-2}
\bibliography{string}


\end{document}


\title{Supplementary Material for \\ ``Statistical Properties of a Fluctuation-Driven Nanomechanical Duffing Resonator"}

\newcommand{\BU}{Department of Mechanical Engineering, Division of Materials Science and Engineering, and the Photonics Center, Boston University, Boston, Massachusetts 02215, USA}

\author{M. Ma}
\affiliation{Department of Mechanical Engineering, Division of Materials Science and Engineering, and the Photonics Center, Boston University, Boston, Massachusetts 02215, USA \looseness=-1}

 \author{N. Welles}
 \affiliation{Department of Mechanical Engineering, Virginia Tech, Blacksburg, Virginia 24061, USA \looseness=-1}

 \author{J. Barbish}
\affiliation{Department of Engineering, James Madison University, Harrisonburg, Virginia 22807, USA \looseness=-1}
 

\author{I. I. Kaya}
\affiliation{SUNUM, Nanotechnology Research and Application Center, Sabanci University, Istanbul, 34956, Turkey \looseness=-1}
\affiliation{Faculty of Engineering and Natural Sciences, Sabanci University, Istanbul, 34956, Turkey \looseness=-1}


\author{M. S. Hanay}
\affiliation{Department of Mechanical Engineering, Bilkent University, Ankara, 06800, Turkey \looseness=-1}
\affiliation{National Nanotechnology Research Center (UNAM), Bilkent University, Ankara, 06800, Turkey \looseness=-1}


\author{O. Svitelskiy}
\affiliation{Department of Physics, Gordon College, Wenham, Massachusetts 01984, USA \looseness=-1}

\author{M. R. Paul}
\email{mrp@vt.edu}
\affiliation{Department of Mechanical Engineering, Virginia Tech, Blacksburg, Virginia 24061, USA \looseness=-1}

\author{K. L. Ekinci}
\email{ekinci@bu.edu}
\affiliation{Department of Mechanical Engineering, Division of Materials Science and Engineering, and the Photonics Center, Boston University, Boston, Massachusetts 02215, USA \looseness=-1}

\date{\today}

\maketitle
\widetext
\tableofcontents

\newpage

\setcounter{equation}{0}
\setcounter{figure}{0}
\setcounter{table}{0}
\setcounter{page}{1}
\makeatletter
\renewcommand{\theequation}{S\arabic{equation}}
\renewcommand{\thefigure}{S\arabic{figure}}
\renewcommand{\thetable}{S\arabic{table}}
\renewcommand{\bibnumfmt}[1]{[S#1]}
\renewcommand{\citenumfont}[1]{S#1}

\maketitle

\section{Experimental Details}
\label{section:SI_Experimental_Details}

\subsection{Measurement of Device Parameters}\label{Sisection:device_param}

\begin{figure}[ht]
    \includegraphics{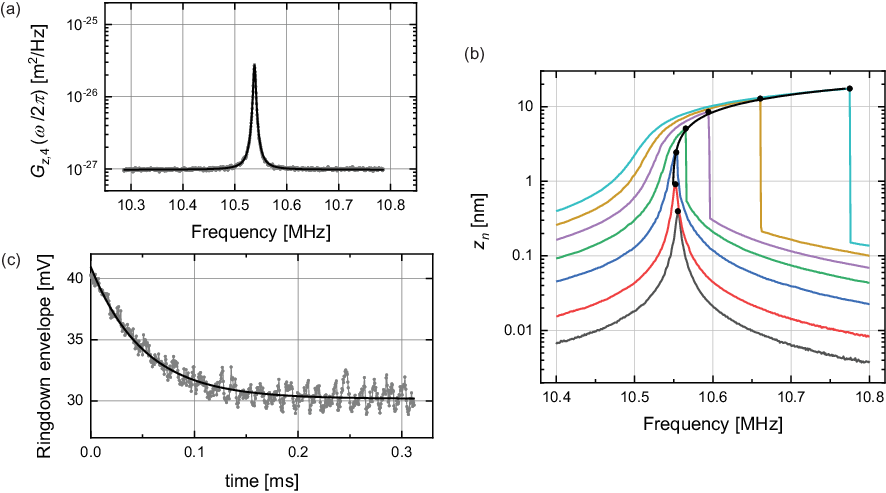}
    \caption{(a) The PSD of a mode under thermal equilibrium \textit{i.e.}, no external drive. A Lorentzian fit is shown with a black continuous line. (b) Amplitude-frequency response curves under harmonic drive near resonant frequency for different drive power levels. The black continuous line is a parabolic fit to the peak amplitudes. (c) The upper envelope of a modal linear ringdown measurement. The exponential decay function with an offset [black continuous line] is fitted to the experimental data [gray symbols].}    
    \label{Figure_S1}
\end{figure}

In this section, we provide details on the measurement of the modal parameters, including the linear spring constant $k_n$, the nonlinear Duffing constant $\alpha_n$, and the quality factor $Q_n$. To extract the experimental value of $k_n$, we turn to the equipartition of energy, where $k_n = \frac{k_B T}{\langle {{z_n}}^2\rangle}$. We use the same notation as in the main text: $k_B$ is the Boltzmann constant, $T = 300\rm~K$ is the ambient temperature, $\langle {{z_n}}^2\rangle$ is the mean-squared amplitude of mode $n$, and is computed as the area under the power spectral density (PSD) measurement in thermal equilibrium (\textit{i,e.,} without external drive). In Fig. \ref{Figure_S1}(a), we show a representative thermal measurement using gray symbols, and a Lorentzian fit to the resonance peak is shown with a black continuous curve.

We measure the Duffing constant $\alpha_n$ using both the noise-based drive shown in the main manuscript and the deterministic drive presented below. Using deterministic frequency sweeps near an eigenmode, we follow the standard steps \cite{landau2013course, nayfeh2008nonlinear,lifshitz2008nonlinear} that lead to the expression for the frequency of the peak $\omega_{p,n}$ as a function of the peak amplitude $z_{p, n}$ and other modal parameters, resulting in the well-studied backbone curve \cite{kozinsky2006tuning, kozinsky2007basins, nayfeh2008nonlinear}:
\begin{equation}\label{eqn:backbone}
    \omega_{p,n} = \omega_n + \frac{3}{8}\left(\frac{{z_{p, n}}^2}{\omega_n}\right)\alpha_n.
\end{equation}
Eq. (\ref{eqn:backbone}) is parabolic in $z_{p, n}$  and is shown as the solid black line in Fig. \ref{Figure_S1}(b). We note that the $z_{p, n}$ corresponds to the peak value of the displacement. We compute the experimental values of $\alpha_n$ by fitting a parabolic function (black continuous line) using measured $z_{p,n}$ values at the largest drives to Eq. (\ref{eqn:backbone}). In the main manuscript, the Duffing constant is determined from stochastic excitation, where the response is based on the mean-squared amplitude $\langle {z_n}^2\rangle$. Accounting for the peak and RMS values in $z_n$, we achieve the same values of $\alpha_n$ using both methods.

We extract $Q_n$ from the ringdown measurements of a mechanical mode driven by a harmonic burst signal in the linear regime close to its resonant frequency $\omega_n$. During a ringdown, the time-dependent modal amplitude ${z_n}(t)$ decays exponentially as
\begin{equation}\label{eqn:ringdown}
    {z_n}(t) = z_{p,n} e^{-{\omega_n t}/2{Q_n}} \cos (\omega_n t+ \phi) + z_{N}.
\end{equation}
Here, $\phi$ is an arbitrary phase, and $z_{N}$ is the technical detection noise that arises in a typical broadband measurement. As shown in Fig. \ref{Figure_S1}(c), $z_{N} \approx 30~\rm mV$, and we also show this detection noise as the red-shaded area in Fig. 3(b, inset) of the main manuscript. From Fig. \ref{Figure_S1}(c), we extract the upper envelope of the ringdown from measurement (gray symbols), \textit{i.e.}, $z_{p,n} e^{-{\omega_n t}/2{Q_n}} + z_{N}$, and find the value of $Q_n$ by fitting an exponential decay function with an offset (black continuous line) to the envelope.

\subsection{Synthetic Drive Noise Properties}


Our synthetic noise is a band-limited white noise generated in MATLAB and continuously played on an arbitrary waveform generator (AWG) to stochastically drive a mechanical mode of our doubly clamped beam resonator. The schematic for the electronic drive is shown in Fig. 1(a) of the main manuscript. The fluctuating component of the noise, $v(t)$, is generated in the time domain and consists of $\sim 10^5$ decaying sine waves. Each sine wave has a randomly assigned frequency, decay time, starting time point, and phase. To ensure noise quality, we set the total number of time points to $N_t = 5\times 10^6$, limited by the AWG's memory depth. We define a uniform probability distribution within a specific range for each wave parameter: the frequency is within the noise bandwidth $\Delta f \approx 1~\rm{MHz}$; the decay time is between $10^{-3}$ and $10^{-2}\rm ~s$; the starting time point is between $0$ and $N_t$; and the starting phase is between $0$ and $2\pi$. Each decaying wave in the time domain is a Lorentzian peak with a finite linewidth in the frequency domain, and a combination of such randomly generated waves results in the band-limited Gaussian noise, shown as the black curve in Fig. 1(c) of the main manuscript.

\begin{figure}[ht]
    \includegraphics{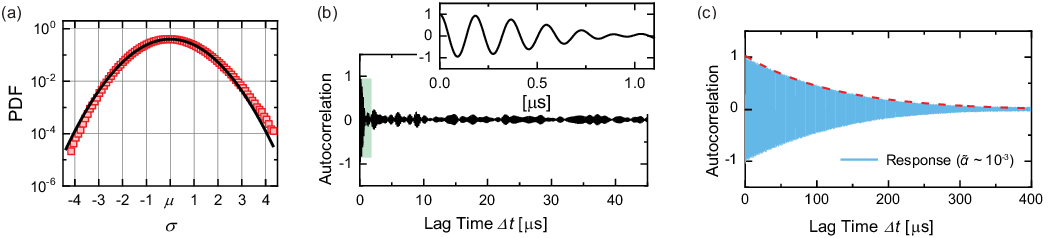}
    \caption{(a) The PDF of the normalized noise power shows a Gaussian distribution. The continuous curve is a Gaussian fit. (b) The autocorrelation function for the fluctuating noise signal $v(t)$. The inset shows the correlation time. (c) Typical autocorrelation of a linear response. The dashed curve is an exponential decay function with $\tau \approx 130~\rm \mu s$.}
    \label{Figure_S15}
\end{figure}

In the main manuscript, we showed that the PDF of the fluctuating noise component $v(t)$ is Gaussian [Fig. 1(b)]. We have also measured the statistics of the drive noise signal we generated. We add a DC offset to our fluctuating noise signal, $V+v(t)$, such that $V\gg v(t)$ and hence $(V+v(t))^2 \approx V^2+ 2Vv(t)$. Since the force generated in the electrothermal actuator is proportional to $(V+v(t))^2$, we compute the PDF of the drive power $(V+v(t))^2$ from measured time traces of $V+v(t)$. The computed PDF is shown in Fig. \ref{Figure_S15}(a) for mode $n=2$, where the $x-$axis is centered at the mean power value $\mu \approx 4\times 10^{-2} ~\rm{V^2} $ and has units of $\sigma  \approx 3.1\times 10^{-3} ~\rm{V^2}$. The noise power has a Gaussian distribution (continuous curve) even for $\gtrsim \pm 3\sigma$. We conclude that any observed deviation from Gaussianity is due to the nonlinear dynamics --- rather than the forcing. 


Finally, we further analyzed the drive noise by recording the signal in the time domain for  $100 ~\rm \mu s$ and computing the normalized autocorrelation function, $\langle v(t)\,v(t+\Delta t)\rangle/\langle v^2\rangle$, shown in Fig. \ref{Figure_S15}(b). The inset of Fig. \ref{Figure_S15}(b) shows that the correlation time is $\sim 1 ~\rm \mu s$, in agreement with $1/\Delta f \approx 1 ~\rm \mu s$, and it takes $\approx 7$ cycles for the autocorrelation function to decay, consistent with $\omega_4/2\pi \Delta f \approx 7.6$. Given that the relaxation time of the driven resonator mode is $2\pi Q_4/\omega_4 \approx 170 ~\rm \mu s $, our force noise can be considered delta-correlated for all practical purposes. We show a typical autocorrelation function of a linear response ($\tilde \alpha \sim 10^{-3}$) in Fig. \ref{Figure_S15}(c) for mode $n=3$. Here, the decay envelope is nicely fitted with an exponential decay function (dashed curve), with the relaxation time constant of $\approx 130~\rm \mu s$. In our experiments, we verified the Gaussianity and the same time and frequency domain properties for the noisy drive for all modes under study.

\subsection{Calibration of $T_{\text{eff}}$}\label{SIsection:Teff_calibration}

\begin{figure}[ht]
    \includegraphics{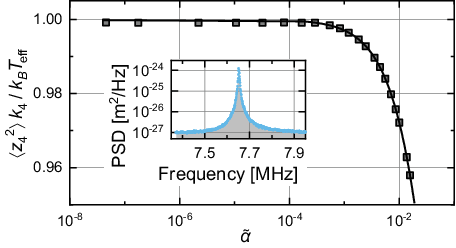}
    \caption{Nondimensional variance as a function of the nonlinear strength parameter $\tilde{\alpha}$. The experimental values of $T_{\text{eff}}$ (symbols) are found from fitting to the theoretical curve. The inset shows a noise-driven response amplitude PSD, where the highlighted area indicates $\langle {{z_4}}^2\rangle$.}    
    \label{Figure_S25}
\end{figure}

We compute the experimental value of $T_{\text{eff}}$ from the theoretical variance of a Duffing resonator, as shown in Eq. (3) in the main manuscript. To better illustrate the deviation of the variance of a Duffing oscillator from the variance of a harmonic oscillator, we show the dimensionless variance ${k_n}\langle {z_n}^2\rangle / k_B T_{\text{eff}}$ for mode $n=4$ in Fig. \ref{Figure_S25}. The theoretical curve for the variance \cite{roberts2003random} as a function of $\tilde{\alpha}$ is shown in Fig. \ref{Figure_S25} as a black continuous curve and is expressed as
\begin{equation}\label{eqn:variance-theory-nondim}
    \frac{k_n\langle {z_n}^2\rangle}{k_B T_{\text{eff}}} =  \pi^{1/2}  \left( \frac{\tilde{\alpha}}{2} \right)^{-1/4} \mathcal{D}_{-3/2}\left( (2\tilde{\alpha})^{-1/2} \right) \mathcal{K}_{1/4}^{-1}\left( (8\tilde{\alpha})^{-1} \right).
\end{equation}

We find the experimental values of $T_{\text{eff}}$ from Eq. (\ref{eqn:variance-theory-nondim}) using the experimental values of $k_n$, $\alpha_n$, $\omega_n$ as described in the main manuscript. We compute the $\langle {{z_n}}^2\rangle$ value by integrating the area under the PSD curves at different power levels. A representative PSD is shown in the inset of Fig. \ref{Figure_S25}, and the area under the PSD curve is highlighted in gray. The experimental $T_{\text{eff}}$ values are shown as symbols. From equipartition of energy, $k_n \langle {{z_n}}^2\rangle/( k_B T_{\text{eff}}) = 1$ for a harmonic resonator. In our experiments, we reach the dimensionless variance values of $\approx 0.96$ due to nonlinear stiffening \cite{barbish2023dynamics}.

\subsection{Calculation of PDF and its Moments from Experimental Data}

We first measure the stochastic response $z_n(t)$ of a noise-driven mode in the time domain at different levels of $\tilde{\alpha}$. For each $z_n(t)$, we create a histogram of $z_n$ with $\sim 50$ discrete interval bins. We normalize the histogram as $\int_{-\infty}^{\infty} P(z_n, \tilde\alpha) \,dz = 1$ to get the corresponding PDF curve, such as the ones shown in Figs. 1(b) and (d) of the main manuscript. For the PDF curves shown in Fig. 4(a) in the main manuscript, we plot the dimensionless displacement, $\tilde{z} = z_n/\sqrt{k_BT_\text{eff}/k_n}$, on the $x$-axis. Using each experimental $P(z_n, \tilde\alpha) \,dz$, we calculate the $m^{\rm{th}}$ order moment of our PDFs, $\langle{z_n}^{m}\rangle$, using the following expression
\begin{equation}\label{SIeqn:moments}
   \langle{z_n}^{m}\rangle = \int_{-\infty}^{\infty}{z_n}^{m}P(z_n, \tilde\alpha)\,dz_n.
\end{equation}

\subsection{Statistical Limits on Observable Tail Events in Finite Gaussian Samples}

As we construct our PDF curves using a finite data set $z_n(t)$ drawn from a nearly Gaussian distribution, the number of standard deviations $n_\sigma$ displayed in our PDF is fundamentally limited by the independent sample size $N$ \cite{westfall2013understanding}. This limitation stems from the extremely fast decay rate of the Gaussian function, which has a quadratic exponential form. To reliably capture a displacement amplitude $|\tilde{z}| > n_\sigma$, we require at least $5$ expected independent sample in the corresponding probability region, \textit{i.e.}, $NP(|\tilde{z}| > n_\sigma) \geq 5$. For a Gaussian PDF,
\begin{equation}\label{SIeqn:n_sigma}
    P(|\tilde{z}| > n_\sigma) \approx \frac{1}{\sqrt{2\pi}} \frac{e^{-{n_\sigma}^{2}/2}}{n_\sigma}.
\end{equation}
In our experiments, the fluctuation amplitude remains coherent for a time span of a ringdown, \textit{i.e.}, $\tau_c \sim 2\pi Q_n/ \omega_n$. As a result, the number of independent response amplitude values is $N\approx t/\tau_c$, where $t$ is the finite time of the measured signal. For the PDF curve shown in Fig. 4(a) of the main manuscript, $N \approx 400$. Following Eq. (\ref{SIeqn:n_sigma}), the maximum number of standard deviations $n_\sigma$ is computed as $n_\sigma \lesssim \sqrt{ 2 \ln \frac{N}{5}} \approx 2.96$. We calculate the $n_\sigma$ values across all modes studied in experiments, and all our PDF curves are accurate up to at least $\approx 2.9|\tilde{z}|$.

\section{Theory}



\subsection{Power Spectral Density (PSD) due to Noise Excitation}\label{SIsection:PSD_theory}

We derive the expressions for the theoretical PSD curves shown in Fig. 2(a) in the main manuscript. We start with the Duffing resonator equation with damping for a single mode $n$,
\begin{equation}\label{eqn:duffing}
    {\ddot{z}_n} + {\frac{{\omega _n}}{{Q_n}}}{\dot{z}_n} + {{\omega_n}^2}{{z_n}} + {\alpha _n}{{z_n}}^3 = \frac{\xi(t)}{m_n}.
\end{equation}
Here, ${z_n}$ denotes the out-of-plane displacement at the modal antinode closest to the center, $\omega_n$ is the angular eigenfrequency, $\alpha_n$ is the nonlinear stiffening coefficient (Duffing constant), $Q_n$ is the quality factor, $\xi(t)$ is the stochastic driving force arising from the input noise, and $m_n$ is the modal mass. 

To obtain the theoretical frequency-response curves for stochastically driven Duffing resonators, we incorporate the method of averages \cite{dykman1971classical,nayfeh2008nonlinear} to remove the fast oscillating component of the dynamics and obtain the slowly varying envelope of the resonator amplitude. Using path integral methods \cite{dykman1971classical,dykman1980time,welles2024mode}, one can solve for the so-called correlator, \textit{i.e.}, the envelope of the autocorrelation function of the response amplitude, as follows:
\begin{equation}\label{eqn:Dykman-correlator}
    \langle A\left( t\right) \bar{A}\left( 0\right)\rangle = \frac{k_B T_{\text{eff}}}{k_n} \frac{e^{{\omega_n t}/{2Q_n}}}{\psi^2\left( t \right)},
\end{equation}
where the overbar denotes complex conjugate, $\psi(t) = \cosh{(at)} + \frac{\omega_n(1-2i\zeta)}{2Q_na}\sinh{(at)}$, $a = \frac{\omega_n}{2Q_n}\sqrt{1 - 4i\zeta}$, and $\zeta = 2Q_n({\omega_{p,n} - \omega_n})/{\omega_n}$. For linear oscillations, $\zeta = 0,$ thus the amplitude decay becomes identical with linear ringdown with the initial amplitude ${k_B T_{\text{eff}}}/{k_n}$ and an envelope of $e^{-\omega_n t/2Q_n}.$ In our experiments, we achieve the highest value of $\zeta \approx 30$. 

Next, we compute the approximate autocorrelation function of a stochastic Duffing resonator by multiplying the correlator from Eq. (\ref{eqn:Dykman-correlator}) with harmonic oscillations at the modal eigenfrequency, \textit{i.e.}, $\langle A\left( t\right) \bar{A}\left( 0\right)\rangle e^{-i\omega_n t}$.
Using the Wiener-Khinchin theorem and the Fourier convention from \cite{paul2006stochastic}, the theoretical PSD of the nonlinear stochastic response is computed as
\begin{equation}\label{eqn:psd-theory}
    G_{z,n}(\omega) = \frac{2}{\pi} \Re \left(\int_{0}^{\infty} \langle A\left( t\right) \bar{A}\left( 0\right)\rangle e^{-i\omega_n t} \,dt \right),
\end{equation}
where $\Re$ corresponds to the real component. Finally, we numerically compute the theoretical response PSD curves for all values of $T_{\text{eff}}$ using Eqs. (\ref{eqn:Dykman-correlator}) - (\ref{eqn:psd-theory}). 

\subsubsection{\textcolor{black}{Calculating the Theory Curves for PSD from Experimental Data}}


\textcolor{black}{In Fig. 2(a) of the main manuscript, we used the following parameters for calculating the theory curves for mode $n=4$: $\omega_n/2\pi \approx 7.66~\rm MHz$, $Q_n \approx 2\times10^3$, $k_n \approx 16~\rm N/m$, and $\alpha_n/4\pi^2 \approx 1.6 \times 10^{-2}~\rm MHz^2/nm^2$. In addition, $T_{\text{eff}}$ values are determined as described in Section~\ref{SIsection:Teff_calibration}, and $\zeta$ values are based on the experimental values of the shifted peak frequencies $\omega_{p,n}$. We also add a small offset $\approx 10^{-27}~\rm m^2/Hz$ to each theoretical PSD curve to account for the detection noise. We use the perturbed experimental values of $\omega_n$, $Q_n$, and $k_n$, where the Joule heating effect from the external drive is accounted for as described in Section~\ref{SectionSI:jouleheating}. For the Duffing constant $\alpha_n$, we use the unperturbed value because $\alpha_n$ is only weakly temperature-dependent, with a temperature dependence of $\lesssim 5\%$ across our experiments.}

\textcolor{black}{In our theoretical PSD curves, the spring constant $k_n$ primarily sets the amplitude value of the peak, while $\alpha_n$ governs the nonlinear broadening of the peak toward higher frequencies. The quality factor $Q_n$ determines the linewidth in the linear regime, but in the nonlinear regime, the broadening is dominated by $\alpha_n$. Consequently, the computed PSD curves are only weakly sensitive to $Q_n$, and varying $Q_n$ within our experimental uncertainty produces a negligible change in the theory curves. We note that the theoretical curves deviate slightly from the experimental data in the nonlinear regime, particularly for the two highest-drive responses shown in Fig. 2(a) of the main text, primarily on the high-frequency side. Because the high-frequency side of the spectrum is particularly sensitive to $\alpha_n$, even a modest shift in $\alpha_n$ can affect the predicted response. Since $\alpha_n$ may vary by up to $\sim5\%$ at the highest drives owing to its weak temperature dependence, we attribute the discrepancy primarily to our use of the unperturbed value of $\alpha_n$ in the calculations.}

\subsection{Displacement PDF and its Moments}
The dimensionless probability distribution function for the displacement of a Duffing resonator is defined with the standard Boltzmann distribution, as shown in Eq. (7) in the main manuscript. To find the precise probability density function (PDF), we find an analytical form for the normalization constant, also known as the partition function~\cite{chandler1987introduction}. 

We find the partition function $\mathcal{Z}(c_1, c_2)$ which normalizes the probability distribution by using the definition of a quartic Gaussian integral~\cite{amdeberhan2012evaluation}. This definition is available in mathematical reference tables (see Eq. 3.469.1 in Gradshteyn and  Ryzhik~\cite{gradshteyn2007table}). This matches the analytical expression we found using Mathematica~\cite{mathematica}. Using the definition from Gradshteyn and Ryzhik, the position component of the partition function is
\begin{equation}\label{eq:quartic-gaussian-integral}
    \mathcal{Z}(c_1, c_2) = \int_{-\infty}^{\infty} \exp(-c_1 z^2 - c_2 z^4) dz = \frac{1}{2} \sqrt{\frac{c_1}{c_2}} \exp\left( \frac{{c_1}^2}{8c_2} \right) \mathcal{K}_{1/4}\left( \frac{{c_1}^2}{8c_2} \right), 
\end{equation}
where $c_1$ and $c_2$ are constants. For our work, we choose $c_1=1/2$, $c_2 = \tilde{\alpha}/4$ based on the nondimensionalization of the nonlinearity to the parameter $\tilde{\alpha}$. This simplifies the partition function to an expression in agreement with Ref.~\cite{roberts2003random}. This reduces the nondimensional PDF to
\begin{equation}\label{eqn:pdf-Z}
    P (\tilde{z},{\tilde{\alpha}}) = \frac{(2\tilde{\alpha})^{-1/2}}{\exp[(8\tilde{\alpha})^{-1}] \mathcal{K}_{1/4}[(8\tilde{\alpha})^{-1}]} \exp\left[\frac{-\tilde{z}^2}{2} \left(1 + \frac{\tilde{\alpha}}{2} {\tilde{z}}^{2}  \right)\right],
\end{equation}
where parameters are defined in the main manuscript. 

To the best of our knowledge, explicit expressions for the higher even-order moments ($m\geq4$) are not explicitly written, but were theorized by Rahman in a footnote~\cite{rahman1996stationary}. We use Mathematica~\cite{mathematica} to take a set of appropriate partial derivatives of Eq. (\ref{eq:quartic-gaussian-integral}) to find the first four even-order moments of the displacement. For example, $\partial \mathcal{Z}(c_1, c_2)/\partial c_1$ from Eq. (\ref{eq:quartic-gaussian-integral}) provides the second order moment $\langle\tilde{z}^2\rangle$. The computational derivatives matched with the recurrence relations of the modified Bessel function of the second kind, see Eq. (9.6.26) from Ref.~\cite{abramowitz1965handbook}. We provide explicit expressions for the even-order moments of a Duffing resonator's response PDF curves $\langle {{z_n}}^{m}\rangle$, for the first four even moments $m=2, 4, 6, 8$. The odd-order moment values vanish due to symmetry in the PDF. The first even moment $m=2$ is the variance of a stiffening Duffing resonator, \textit{i.e.},
\begin{equation}\label{SIeqn:z2}
    \langle {\tilde{z}}^2\rangle = \frac{1}{2\tilde{\alpha}} \left( \frac{\mathcal{K}_{3/4}((8\tilde{\alpha})^{-1})}{\mathcal{K}_{1/4}((8\tilde{\alpha})^{-1})} -1 \right),
\end{equation}
which is equivalent to the variance expression from Roberts and Spanos~\cite{roberts2003random}. The higher-order moments are expressed as follows:
\begin{equation}\label{SIeqn:z4}
    \langle {\tilde{z}}^4\rangle = \frac{1}{2\tilde{\alpha}^2} \left(1 + 2\tilde{\alpha} - \frac{\mathcal{K}_{3/4}((8\tilde{\alpha})^{-1})}{\mathcal{K}_{1/4}((8\tilde{\alpha})^{-1})} \right),
\end{equation}

\begin{equation}\label{SIeqn:z6}
    \langle {\tilde{z}}^6\rangle = \frac{(1 + 15\tilde{\alpha} + 60\tilde{\alpha}^2)\mathcal{K}_{3/4}((8\tilde{\alpha})^{-1})-(1 + 5\tilde{\alpha})\mathcal{K}_{7/4}((8\tilde{\alpha})^{-1})}{2\tilde{\alpha}^3 \mathcal{K}_{1/4}((8\tilde{\alpha})^{-1})},
\end{equation}

\begin{equation}\label{SIeqn:z8}
    \langle {\tilde{z}}^8\rangle = \frac{(1 + 10\tilde{\alpha}(1+\tilde{\alpha}))\mathcal{K}_{7/4}((8\tilde{\alpha})^{-1}) - (1 + 20\tilde{\alpha} (1+6\tilde{\alpha}(1+\tilde{\alpha})))\mathcal{K}_{3/4}((8\tilde{\alpha})^{-1})}{2\tilde{\alpha}^4 \mathcal{K}_{1/4}((8\tilde{\alpha})^{-1})}.
\end{equation}
Using Eqs. (\ref{SIeqn:z2}) - (\ref{SIeqn:z8}), we compute the standardized moment values ${\langle{z_n}^{m}\rangle}/{\langle{z_n}^{2}\rangle^{m/2}}$ as a function of $\tilde{\alpha}$. We show the computed theory curves in Fig. 4(c) of the main manuscript.

\section{Error Sources} \label{SectionSI:errors}

\subsection{Joule Heating due to Noise Drive} \label{SectionSI:jouleheating}

\begin{figure}[ht]
    \includegraphics{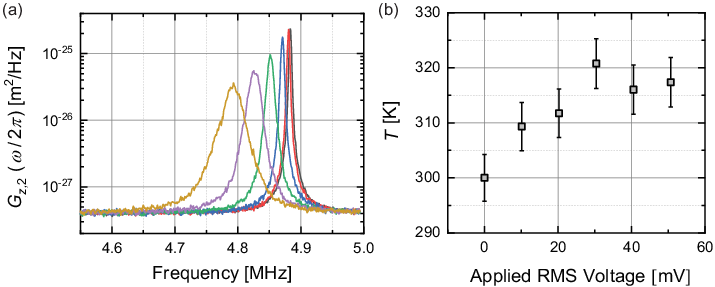}
    \caption{(a) PSD curves as the rms voltage increases. (b) Estimated thermodynamic temperature as a function of applied rms voltage. }    
    \label{Figure_S4}
\end{figure}

The thermodynamic temperature $T$ in our system increases due to Joule heating from the electrothermal noise drive. To estimate $T$, we use the following protocol: we first measure all the mechanical parameters without any heating or drive as described above in Section \ref{Sisection:device_param} and establish baselines. We then apply a known \textit{off-resonance} electrical voltage comparable to the drive voltages and measure the fluctuations of the mode again. The effect of this electrical voltage is to heat the beam, without applying mechanical force to the mode. A typical measurement is shown in Fig. \ref{Figure_S4}(a). The perturbed eigenfrequency $\omega_n$ is immediately available from the downshifted peak; the perturbed spring constant $k_n$ is found from $\omega_n$ by assuming the mode mass does not change; the quality factor $Q_n$ can be found from the linewidth. To find $T$, we assume that the beam is still close to equilibrium since it is not driven by a force noise --- even though heat is dissipated in the system. We integrate the area under the curve and find $\langle{z_n}^2\rangle$.  Then, using the perturbed $k_n$ values, we estimate the temperature increase from the equipartition of energy, \textit{i.e.}, $T = k_n \langle{z_n}^2\rangle/k_B$. The values of $T$ found using this protocol are shown in Fig. \ref{Figure_S4}(b) as a function of the applied rms voltage value.

Using this approach, we estimate the maximum temperature increase of each mode for the maximum applied drive power, as shown in Table \ref{SItable1:parameters}. Also shown are the values of the mechanical parameters at this maximum temperature. We also confirm that, as a result of Joule heating, the quality factors $Q_n$ decrease by a factor $\lesssim 2-3$ compared to the ringdown $Q_n$ values shown in Table I of the main manuscript. In this work, the decrease in $Q_n$ does not strongly affect the results, such as the fits discussed in Section \ref{SIsection:PSD_theory}. We therefore used the unperturbed $Q_n$, where needed.


\begin{table}[htbp]
\caption{Mechanical parameters for the first 4 modes under thermal equilibrium and DC offset $V$.}\label{SItable1:parameters}
\renewcommand{\arraystretch}{1.3}
\newcolumntype{Y}{>{\centering\arraybackslash}X}
\begin{tabularx}{\textwidth}{Y|Y|YY|YY}
 \hline
  & &\multicolumn{2}{c|}{Thermal Equilibrium}& \multicolumn{2}{c}{Maximum Drive} \\\cline{3-6}
 Mode $n$&Maximum $T$& $\omega_n/2\pi $  & $k_n$ & $\omega_n/2\pi$ & $k_n$ \\
 &$\rm (K)$& $(\rm{MHz})$  & $(\rm N/m)$& $(\rm{MHz})$ & $(\rm N/m)$\\

 \hline
1 & $350 \pm 10$& $2.37 $& $0.91$& $1.74$&$0.78$\\
2 & $380 \pm 20$& $4.90$  & $3.44$& $2.75$&$2.50$\\
3 & $350 \pm 10$ & $8.01$  & $10.0$& $5.37$& $8.18$\\
4 &  $400 \pm 20$ & $10.5$    & $19.2$ & $7.66$&$16.3$\\
\hline
\end{tabularx}
\end{table}



\subsection{Transducer Nonlinearity}


We detect the stochastic amplitude fluctuations using a homodyne Michelson interferometer. Here, the mechanical displacement is transduced into a phase difference between the reference and object laser beams, resulting in a measurable change in optical intensity on the photodetector. We use the following expression to convert the detected optical intensity $D$ to the displacement $z_n$ \cite{wagner1990optical, dolleman2017amplitude}
\begin{equation}\label{SIeqn:intereferometer-nonlinear}
    D = ({a_o}^2 + {a_r}^2)\left[ 1 + 2 \frac{a_r a_o}{{a_o}^2 + {a_r}^2}\cos\left(k (d_r - d_o) + 2kz_n\right) \right].
\end{equation}
Here, $a_o$ and $a_r$ are the electric field amplitudes of the object and the reference laser beams, respectively, $d_r - d_o$ is the path difference between the two laser beams, and is set to $d_r - d_o = \lambda/4$ for highest sensitivity \cite{wagner1990optical}, $\lambda = 633~\rm nm$ is the laser wavelength, and $k = 2\pi/\lambda$ is the wavenumber. In our experiments, the highest measured peak displacement is ${z_n}\approx 45~\rm nm$, which results in $\approx 5\%$ error with respect to the linearized Eq. (\ref{SIeqn:intereferometer-nonlinear}). To this end, we created a numerical calibration curve to convert the measured $D$ to $z_n$ based on Eq. (\ref{SIeqn:intereferometer-nonlinear}) to correct for the slight nonlinearity in the optical transducer.




\subsection{Effect of Noise Force Bandwidth}

\begin{figure*}[h]
\centering
    \includegraphics{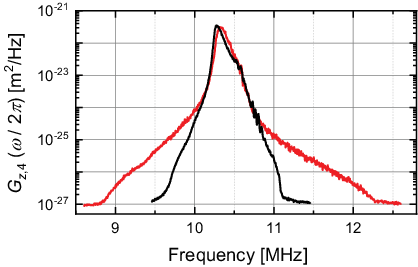}
    \caption{Noise-driven PSD response curves for mode $n=4$ on semilogarithmic axes. The noise drive bandwidth is $\approx 1.6 \rm ~MHz$ (black trace) and $\approx 3.3 \rm ~MHz$ (red trace).}
    \label{R1Fig:BW}
\end{figure*}

We have conducted experiments in which we increased the noise bandwidth to demonstrate its effect on the response.  In Fig. \ref{R1Fig:BW}, we show two response PSD curves, $G_{z,4}({\omega\over 2\pi})$, driven by noise with a bandwidth of $\approx 1.6 \rm ~MHz$, which is the bandwidth used in our experiments (black trace) and $\approx 3.3 \rm ~MHz$ (red trace) centered around the peak. The response here is nonlinear, and $T_{\text{eff}} \sim 10^7~\rm K$. While a larger bandwidth results in broader tails in the response, the response at the tails is 2-3 orders of magnitude smaller and negligible.  When we integrate the areas under these two response curves, we find that the mean-square amplitude $\langle {z_4}^2\rangle$ increases by $\lesssim 4\%$ for the wider bandwidth.

\bibliographystyle{apsrev4-1}
\bibliography{string}